\pdfoutput=1
\documentclass[twocolumn,english,aps,prx,superscriptaddress,bibnotes,amsmath,amssymb]{revtex4-1}

\usepackage{amsmath}
\usepackage{changes}
\usepackage{graphicx}
\usepackage{hyperref}
\usepackage{changes}

\begin{document}

\title{Mid-infrared frequency comb generation with silicon nitride nano-photonic waveguides}

\author{Clemens~Herkommer}
\thanks{These authors contributed equally to this work}
\affiliation{{\'E}cole Polytechnique F{\'e}d{\'e}rale de Lausanne (EPFL), LPQM, CH-1015 Lausanne, Switzerland}
\affiliation{Technische Universit{\"a}t M{\"u}nchen (TUM), Physik-Department, D-80333 M{\"u}nchen, Germany}

\author{Adrien~Billat}
\thanks{These authors contributed equally to this work}
\affiliation{{\'E}cole Polytechnique F{\'e}d{\'e}rale de Lausanne (EPFL), PHOSL, CH-1015 Lausanne, Switzerland}

\author{Hairun~Guo}
\thanks{These authors contributed equally to this work}
\affiliation{{\'E}cole Polytechnique F{\'e}d{\'e}rale de Lausanne (EPFL), LPQM, CH-1015 Lausanne, Switzerland}

\author{Davide~Grassani}
\affiliation{{\'E}cole Polytechnique F{\'e}d{\'e}rale de Lausanne (EPFL), PHOSL, CH-1015 Lausanne, Switzerland}

\author{Chuankun~Zhang}
\affiliation{{\'E}cole Polytechnique F{\'e}d{\'e}rale de Lausanne (EPFL), LPQM, CH-1015 Lausanne, Switzerland}
\affiliation{Tsinghua University, Beijing 100084, China}

\author{Martin~H.~P.~Pfeiffer}
\affiliation{{\'E}cole Polytechnique F{\'e}d{\'e}rale de Lausanne (EPFL), LPQM, CH-1015 Lausanne, Switzerland}

\author{Camille-Sophie~Br\`es}
\email{camille.bres@epfl.ch}
\affiliation{{\'E}cole Polytechnique F{\'e}d{\'e}rale de Lausanne (EPFL), PHOSL, CH-1015 Lausanne, Switzerland}

\author{Tobias~J.~Kippenberg}
\email{tobias.kippenberg@epfl.ch}
\affiliation{{\'E}cole Polytechnique F{\'e}d{\'e}rale de Lausanne (EPFL), LPQM, CH-1015 Lausanne, Switzerland}

\begin{abstract}
  Mid-infrared optical frequency combs are of significant interest for molecular spectroscopy due to the large absorption of molecular vibrational modes on one hand, and the ability to implement superior comb-based spectroscopic modalities with increased speed, sensitivity and precision on the other hand. Substantial advances in mid-infrared frequency comb generation have been made in recent years based on nonlinear frequency conversion, microresonator Kerr frequency combs, quantum cascade lasers and mode locking regimes. Here we demonstrate a simple, yet effective method for the direct generation of mid-infrared optical frequency combs in the region from ${2.5-4~\mu{\rm m}}$, i.e. ${2500-4000~{\rm cm}^{-1}}$ covering a large fraction of the functional group region, directly from a conventional and compact erbium-fiber-based femtosecond laser in the telecommunication band (i.e. ${1.55~\mu{\rm m}}$). The wavelength conversion is based on dispersive wave generation within the supercontinuum process in large-cross-section and dispersion-engineered silicon nitride (${\rm Si_3N_4}$) waveguides. The long-wavelength dispersive wave, with its position lithographically determined, performs as a mid-infrared frequency comb, whose coherence is demonstrated via optical heterodyne measurements. Such a simple and versatile approach to mid-infrared frequency comb generation is suitable for spectroscopic applications in the first mid-infrared atmospheric window. Moreover, the compactness and simplicity of the approach have the potential to realize compact dual-comb spectrometers. The generated combs have a fine teeth-spacing, making them also suitable for gas phase analysis.
\end{abstract}

\maketitle

\section{Introduction}
\label{sec:intro}

Mid-infrared (Mid-IR) optical frequency combs \cite{schliesser2012mid} are of significant interest for molecular spectroscopy, due to the fact that many molecules have strong vibrational transitions in this spectral region. In addition, optical frequency combs can constitute superior spectroscopic modalities, such as dual-comb spectroscopy \cite{lee2001ultrahigh, schiller2002spectrometry, keilmann2004time, schliesser2005frequency, yasui2005asynchronous, coddington2008coherent, giaccari2008active, coddington2010coherent, bernhardt2010cavity, zhang2013mid, ideguchi2014adaptive, villares2014dual, coddington2016dual, millot2016frequency, suh2016microresonator, yu2016silicon, link2017dual} which enables rapid, broadband and precise analysis, or ultra-sensitive cavity-enhanced spectroscopy \cite{thorpe2006broadband, bernhardt2010cavity, bjork2016direct}.
Triggered by a significant number of applications such as pharmaceutical, environmental or medical breath analysis \cite{manolis1983diagnostic, miekisch2004diagnostic, mccurdy2007recent, thorpe2008cavity, wang2009breath, wojtas2014cavity}, mid-IR spectroscopy has attracted substantial attention in the past decade.
A variety of methods have been developed to generate mid-IR frequency combs \cite{schliesser2012mid}, such as nonlinear frequency conversion including both optical parametric oscillators (OPOs) \cite{adler2009phase, leindecker2011broadband, leindecker2012octave, vodopyanov2011mid, petrov2012parametric, jin2014two} and difference frequency generation (DFG) \cite{keilmann2012mid, ruehl2012widely, cruz2015mid}, phase-locked states in quantum cascade lasers \cite{hugi2012mid}, microresonator based Kerr frequency combs \cite{wang2013mid, griffith2015silicon, luke2015broadband, yu2016mode} or using mid-IR mode-locked lasers \cite{sorokin2007sensitive, vasilyev2014kerr, vasilyev2016multi}.

Moreover, recent work has also highlighted the potential to generate mid-IR optical frequency combs from coherent supercontinuum process in nano-photonic waveguides \cite{lee2014midinfrared, kuyken2015octave}. Supercontinuum generation (SCG) is an effective method of producing ultra-broadband spectra from mode-locked pulsed lasers, which has been well studied in photonic crystal fibers \cite{dudley2006supercontinuum}. For sufficiently short pulses and in a short interaction length, the spectrum can be coherent, thereby preserving the comb structure of the seed laser, which subsequently enables the carrier-envelope phase stabilization via self-referencing \cite{jones2000carrier, cundiff2003colloquium}.
When applying this process in chip-based nano-photonic waveguides, the pulse energy can be substantially reduced owing to the large Kerr nonlinearity and tight confinement. Today, nano-photonic waveguide-based SCG has been extensively studied in the visible and near-infrared (near-IR), particularly those made from silicon and silicon nitride (${\rm Si_3N_4}$) \cite{Halir2012, Epping2015, Zhao2015, ChavezBoggio2016, Liu2016, Porcel2017}, and has enabled carrier-envelope frequency detection \cite{mayer2015frequency, yoonoh2017coherent, hickstein2017ultrabroadband}, full phase stabilization of frequency combs \cite{carlson2017photonic, carlson2017self}, as well as offset-free frequency comb generation by DFG \cite{mayer2016offset}.
Yet, accessing the mid-IR region and generating coherent frequency combs remain a challenge. While mid-IR frequency comb generation from coherent SCG has been reported in silicon waveguides \cite{kuyken2015octave}, it required using pump lasers placed at the onset of the mid-IR (${\sim 2.5 ~\mu{\rm m}}$) to avoid silicon two-photon absorptions (TPA). For accessing the mid-IR, dispersion engineering actually plays a key role in extending the long wavelength portion of the supercontinuum \cite{Lau2014mirSCG}.
Soliton induced dispersive wave generation is the mechanism \cite{Akhmediev1995} that can induce a coherent and efficient light conversion over a large frequency span, and facilitate ultra-broadband SCG beyond the self-phase modulation (SPM) regime.
In this context, the emergence of a zero dispersion wavelength at the longer wavelength side to the pump source is necessarily required \cite{frosz2005role, lee2014midinfrared, Lau2014mirSCG}, which however is also challenged by the fact that the waveguide dispersion has to counterbalance strong and anomalous material dispersion, implying an operation approaching the waveguide cut-off region. Although the cut-off is not present for fully clad waveguides, the loss of confinement towards the mid-IR causes the cladding absorption to become relevant.

\begin{figure*}[t]
  \centering{
  \includegraphics{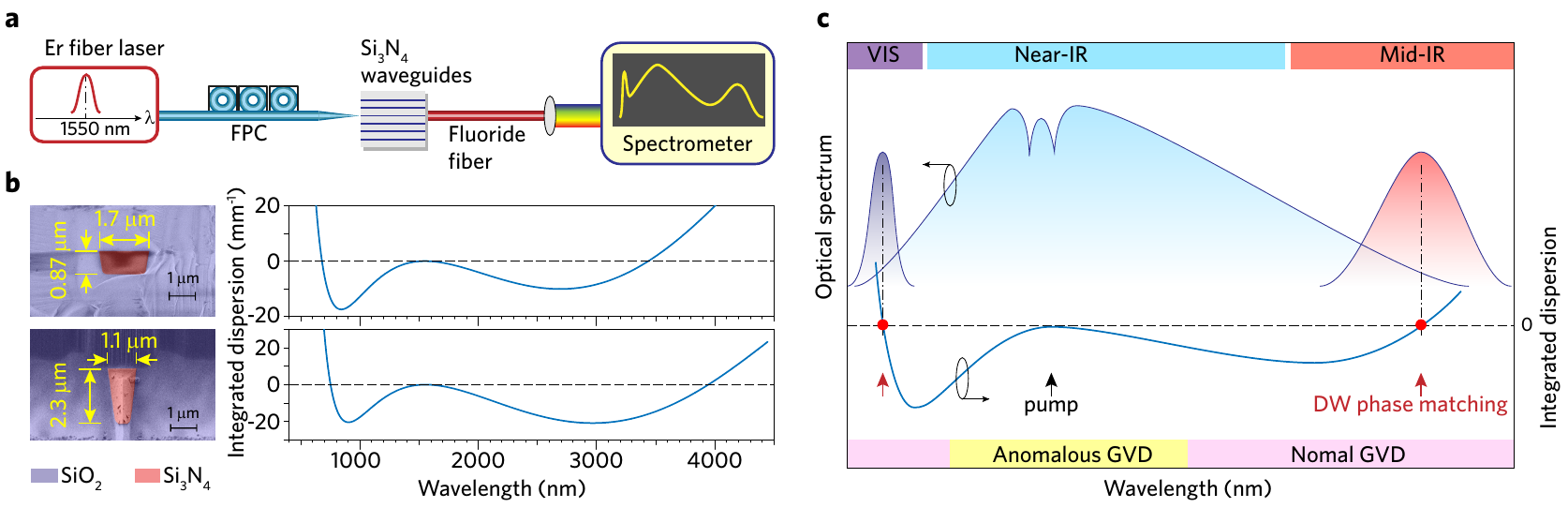}
  \caption{\label{fig-1} \textbf{Principle of mid-IR frequency comb based on dispersive wave generation.} (a) Experimental setup in which the pump is a commercial erbium-fiber based femtosecond pulsed laser centered at ${1.55 ~\mu{\rm m}}$ (ELMO, Menlo System GmbH). The spectrum is measured by both an optical spectral analyzer (OSA) and an Fourier-transform infrared (FTIR) spectrometer. FPC: fiber polarization controller. (b) Scanning electron microscope (SEM) pictures of the cross sections of two classes of ${\rm Si_3N_4}$ waveguides fabricated using the photonic Damascene process: the first class contains waveguides with a core height of ${0.87 ~\mu{\rm m}}$ while the second type has a large cross section with a core height of ${2.3 ~\mu{\rm m}}$. The integrated dispersion of both waveguides are presented, which are calculated by a finite-element method (COMSOL). (c) Schematic representation of soliton induced dispersive wave (DW) generation: by means of dispersion engineering producing higher-order dispersion, dispersive wave is phase-matched to the dispersionless soliton pulse seeded by the pump (where integrated dispersion equals zeros). This leads to the generation of dispersive waves in both the visible and the mid-infrared. While the soliton is supported with anomalous group velocity dispersion (GVD), dispersive waves are in the normal GVD regime.}
  }
\end{figure*}

Here we show that by using ${\rm Si_3N_4}$ nano-photonic waveguides these challenges can be overcome.
We demonstrate the ability to synthesize mid-IR frequency combs directly from an erbium-fiber based femtosecond laser at ${1.55 ~\mu{\rm m}}$ in the telecommunication band (telecom-band), based on mid-IR dispersive wave emission.
${\rm Si_3N_4}$ waveguide is known for its large band-gap and absence of TPA in the telecom-band, as well as large effective nonlinearity and wide transparency window ranging from the visible to the mid-IR (${0.4-4.5 ~\mu{\rm m}}$) \cite{moss2013new}. We developed a dedicated large-cross-section ${\rm Si_3N_4}$ waveguide that supports both mid-IR wave confinement and flexible dispersion engineering.
We demonstrate that the generated mid-IR comb inherits the high level of phase coherence from the pump laser, and with a lithographic control such a comb can cover up to ${\rm 4 ~\mu{\rm m}}$, thereby covering the functional group region including common single-bond vibrational modes such as ${\rm C-H}$, ${\rm O-H}$, and ${\rm N-H}$.

\section{Setup and principle}
\label{sec:setup}

\begin{figure*}[t]
  \centering{
  \includegraphics{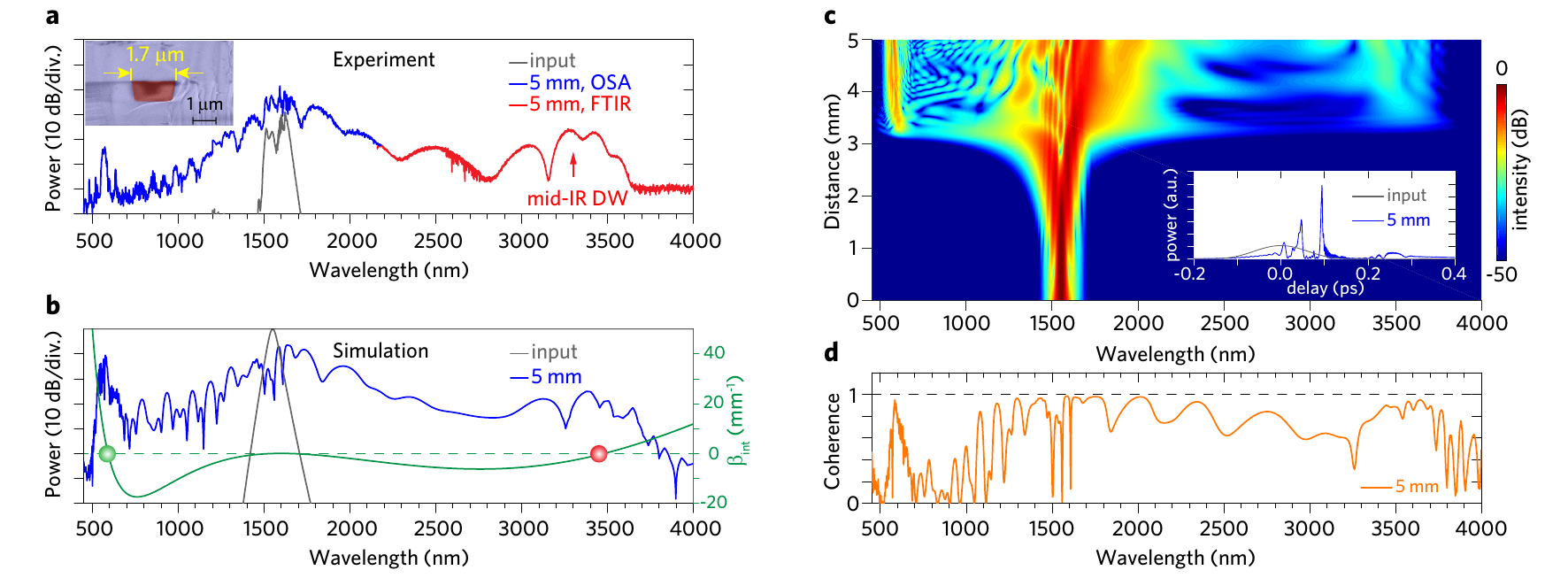}
  \caption{\label{fig-2} \textbf{Comparison between supercontinuum generation experiment and numerical simulation.} (a) Experimentally observed supercontinuum in a waveguide with ${1.7~\mu{\rm m}}$ width and ${0.87~\mu{\rm m}}$ height, measured with a combination of OSA (blue line) and FTIR (red line). The experimental input pump spectrum is shown in grey. (b) Simulated supercontinuum (blue line) in the same waveguide, in excellent agreement with (a). The soliton number is ${N = 5.8}$. The input pulse (grey line) has a duration of ${65 ~{\rm fs}}$, and a ${5 ~{\rm kW}}$ peak power. To account for the dispersion accumulated in the fiber components before the waveguide, a frequency chirping ${e^{-i \mathcal{C} ( \omega - \omega_p )^2} }$, ${\mathcal{C}=1300 ~{\frac{\rm fs^2}{2\pi}}}$, is applied to the input pulse.
  (c) Simulated pulse spectral evolution over a propagation distance of ${5~\rm{mm}}$. Inset: normalized temporal pulse envelope at the waveguide output (blue line) which exhibits the soliton fission effect (soliton split into three components). The input pulse envelope is shown in grey. (d) Modulus of the first order degree of coherence at the waveguide output, showing a high level of coherence particularly for the mid-IR dispersive wave. The coherence simulation is performed with a ${10~{\rm GHz}}$ frequency resolution and with one-photon-per-mode noise, which is appropriate for the case of  a ${100 ~{\rm MHz}}$ pulse repetition rate \cite{Dudley2}.}
  }
\end{figure*}

The experimental setup is shown in Fig. \ref{fig-1}(a). The pump source is a compact mode-locked, erbium-fiber based femtosecond laser (ELMO, Menlo Systems GmbH) that emits a pulse train with a repetition rate of ${\sim 100 ~{\rm MHz}}$. The laser pulse duration (full width at half maximum, FWHM) is ${<90 ~{\rm fs}}$, average power ${>110 ~{\rm mW}}$, pulse energy ${\sim 1 ~{\rm nJ}}$. The laser is connected to a fiber polarization controller (FPC) and coupled into the chip via a lensed fiber, and with inverse tapers at both input and output sides of the ${\rm Si_3N_4}$ waveguide. The insertion loss is ${< 3 ~{\rm dB}}$. It is noted that the pump pulse duration is broadened to ${\sim 130 ~{\rm fs}}$ before the lensed fiber, as measured by an autocorrelator, due to the dispersive effect induced by the FPC. Approximately, a pulse energy of ${\sim 0.6 ~{\rm nJ}}$ is launched into the waveguide.  The waveguide output is then collected by a multimode fluoride fiber (NA ${\sim 0.26}$). The spectra are recorded with two grating-based optical spectrum analyzers (OSAs) covering the wavelength range from ${0.35 - 2.4 ~\mu{\rm m}}$, and with a Fourier-transform infrared (FTIR) spectrometer resolving wavelengths up to ${5 ~\mu{\rm m}}$.

We investigated two classes of ${\rm Si_3N_4}$ waveguides, see Fig. \ref{fig-1}(b), both are fabricated using the photonic Damascene process \cite{Pfeiffer2016} which allows for a deposited ${\rm Si_3N_4}$ layer as thick as ${\sim 1 ~\mu{\rm m}}$. While the first class of waveguides has a ${\rm Si_3N_4}$ core height of ${0.87 ~\mu{\rm m}}$, the second class has a much larger waveguide cross-section making use of the conformal deposition of ${\rm Si_3N_4}$ in LPCVD (cf. \emph{Method}). This leads to a core height as large as ${2.3 ~\mu{\rm m}}$ while the core width is limited to ${2.0 ~\mu{\rm m}}$ (i.e twice the deposition thickness). The waveguide length is ${5 ~{\rm mm}}$.

The design of the waveguide is mainly reflected on lithographic control of the ${\rm Si_3N_4}$ core width.
By reaching a compromise between mid-IR wave confinement and dispersion engineering, the waveguide would enable both the soliton regime on the pump pulse and the mid-IR dispersive wave generation.
Physically, the soliton regime requires anomalous group velocity dispersion (GVD) in the waveguide such that soliton pulses are supported as a result of the balance between the dispersion and the nonlinearity.
For coherent SCG, the waveguide is designed for the nonlinear effect to be stronger than the dispersion effect on the pump pulse, namely a dispersion length exceeding the nonlinearity length, i.e. ${L_D > L_N}$ (${L_D = \frac{T^2}{|\beta^{(2)}|}}$, ${T}$ is the pump pulse duration, ${\beta^{(2)}}$ is the GVD, ${L_N = (\gamma P)^{-1}}$, ${\gamma = \frac{\omega n_2}{c A_{\rm eff}}}$ is the effective nonlinearity, ${n_2}$ is the material nonlinear refractive index, ${A_{\rm eff}}$ is the nonlinear effective mode area, ${P}$ is the peak power of the pump pulse).
This enables a soliton number (${N^2 = \frac{L_D}{L_N}}$) greater than unity such that, during the pulse propagation, soliton self-compression occurs and the spectrum is broadened via SPM \cite{agrawal2013nonlinear}.
Theoretically, the first compression point ${L_c}$, corresponding to the first occurrence of the SCG, is scaled by the soliton number (${L_c \propto \frac{L_D}{N}}$), whereas the compression factor revealing the capability of SPM-induced broadening is ${\frac{T}{T_c}\propto N}$ (${T_c}$ is the pulse duration at the first compression point) \cite{kelley2002scaling}.

SPM is a coherent nonlinear mechanism when the soliton number is relatively low, typically ${1 \le N \le 10}$. Therefore, the determination of the soliton number (${N}$) indicates a trade-off between the spectral coherence and the bandwidth of the supercontinuum.
In addition to SPM, higher-order dispersion will induce perturbations to the soliton dynamics, in particular leading to the dispersive wave generation that can further extend the bandwidth of the soliton-based supercontinuum.
This enables octave-spanning SCG while maintaining a high level of coherence.
In ${\rm Si_3N_4}$ waveguides, coherent SCG in combination with dispersive wave generation can be implemented within a distance of a few millimeters when having a pump of femtosecond pulses with kilo-watt peak power \cite{Epping2015,Zhao2015,mayer2015frequency,mayer2016offset,Liu2016,Porcel2017,Carlson2017NIST}.
Importantly, the phase matching condition between the dispersive wave (at a frequency ${\omega_d}$) and the \emph{dispersionless} soliton pulse (centered at frequency ${\omega _s}$ and having a group velocity ${v_g}$) is:
\begin{equation}
\beta(\omega _d) = \beta(\omega _s) + (\omega _d - \omega _s) \cdot v_g^{-1} + \frac{\gamma P}{2}
\label{equ-pm}
\end{equation}
where ${\beta(\omega)}$ is the propagation constant of the wave propagating in the waveguide.
Since the nonlinear contribution to the soliton phase (${\gamma P}$) is negligible compared to ${\beta(\omega)}$, equation (\ref{equ-pm}) can be rewritten to express the phase mismatch between the dispersive wave and the soliton as:
\begin{align}
\Delta \beta(\omega) & \approx \beta(\omega) - \beta(\omega _s) - (\omega - \omega _s) \cdot v_g^{-1} \\
& =\sum_{m=2}{\frac{(\omega-\omega_s)^m}{m!} \cdot \frac{d^m}{d\omega}\beta(\omega_s)} \buildrel \Delta \over = \beta_{int}
\label{equ-dint}
\end{align}
which is equivalent to the integrated dispersion (defined as ${\beta_{int}}$) in the waveguide.
In addition, ${\rm Si_3N_4}$ has a negligible Raman effect compared to silica (as recently measured \cite{Karpov2016}), which explains the absence of Raman-induced soliton fission as well as the Raman self-frequency shift in the SCG.
Based on a full knowledge of material properties (e.g. the refractive index of ${\rm Si_3N_4}$ and the Kerr nonlinearity) and a precise lithographic control of the geometry in the nano-fabrication, the waveguide dispersion as well as the nonlinear effective mode area can be simulated by a finite element method (COMSOL).

\begin{figure*}[t]
  \centering{
  \includegraphics{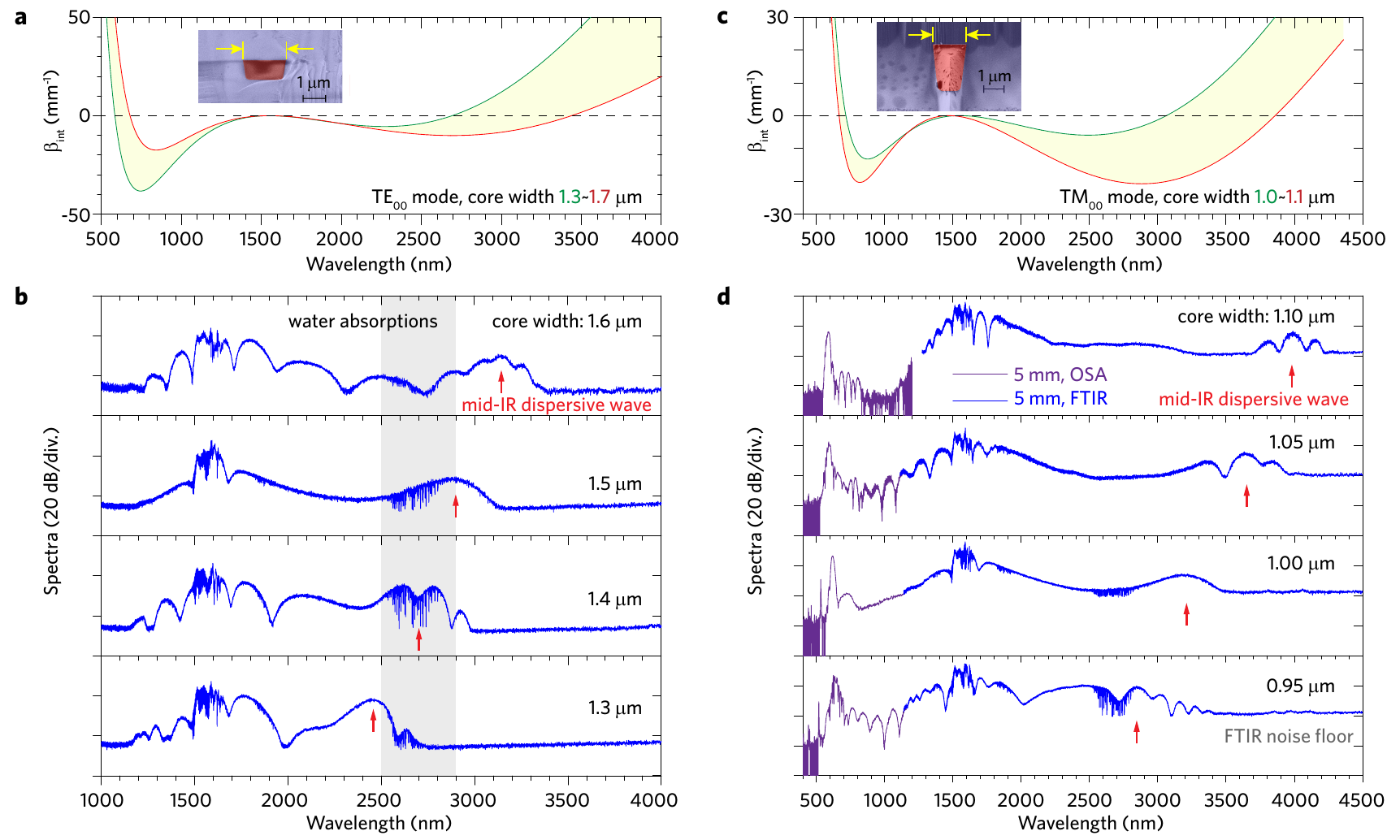}
  \caption{\label{fig-3} \textbf{Lithographical tuning of the mid-IR frequency comb.} (a) Simulated integrated dispersion for the ${0.87 ~\mu{\rm m}}$ thick ${\rm Si_3N_4}$ waveguides. The considered core width is between ${1.3~\mu{\rm m}}$ (green line) and ${1.7~\mu{\rm m}}$ (red line). (b) Spectra generated when pumping such  waveguides (length ${5 ~{\rm mm}}$) with the erbium-fiber based femtosecond laser (pulse energy at the facet ${\sim 0.6 ~{\rm nJ}}$). The conversion efficiency into the mid-IR (${> 2.5 ~\mu{\rm m}}$) is ca. ${1-10 ~\%}$. (c) Simulated integrated dispersion for the ${2.3~\mu{\rm m}}$-thick waveguides. The width values are between ${1.0~\mu{\rm m}}$ (green line) and ${1.1~\mu{\rm m}}$ (red line). (d) Supercontinuum generation with mid-IR dispersive waves in such large-size waveguides. Spectra are recorded using a grating OSA (dark blue line) and an FTIR (blue line). For the spectra approaching ${4.0~\mu{\rm m}}$ low-power spectral regions are not resolved due to the noise floor of the FTIR.}
  }
\end{figure*}

Profiles of the integrated dispersion corresponding to both classes of ${\rm Si_3N_4}$ waveguides are shown in Fig. \ref{fig-1}(b), in which a dispersive wave phase matching point can be found in the mid-IR range.
In particular, with a large flexibility of dispersion engineering enabled in large-cross-section waveguides, such phase matching is further extended to ${4 ~\mu{\rm m}}$.
Therefore, mid-IR dispersive wave generation is expected in such waveguides, as schematically indicated in Fig. \ref{fig-1}(c).

\section{Mid-infrared frequency comb based on dispersive wave generation}
\label{sec:scg}

The first experiment is carried out in a set of waveguides with a height of ${0.87~\mu{\rm m}}$ (identical to the thickness of the deposited ${\rm Si_3N_4}$ layer) and with a waveguide width of ${1.3-1.7~\mu{\rm m}}$, where octave-spanning supercontinua are observed.
In particular, in the waveguide with the width of ${1.7~\mu{\rm m}}$, the spectrum covers more than two octaves, see Fig. \ref{fig-2} (a), spanning from the visible (${0.56 ~\mu{\rm m}}$) to the mid-IR (${3.6 ~\mu{\rm m}}$), and containing two dispersive waves at both edges of the spectrum.
Light propagates in the ${\rm TE_{00}}$ mode.

\begin{figure*}[t]
  \centering{
  \includegraphics{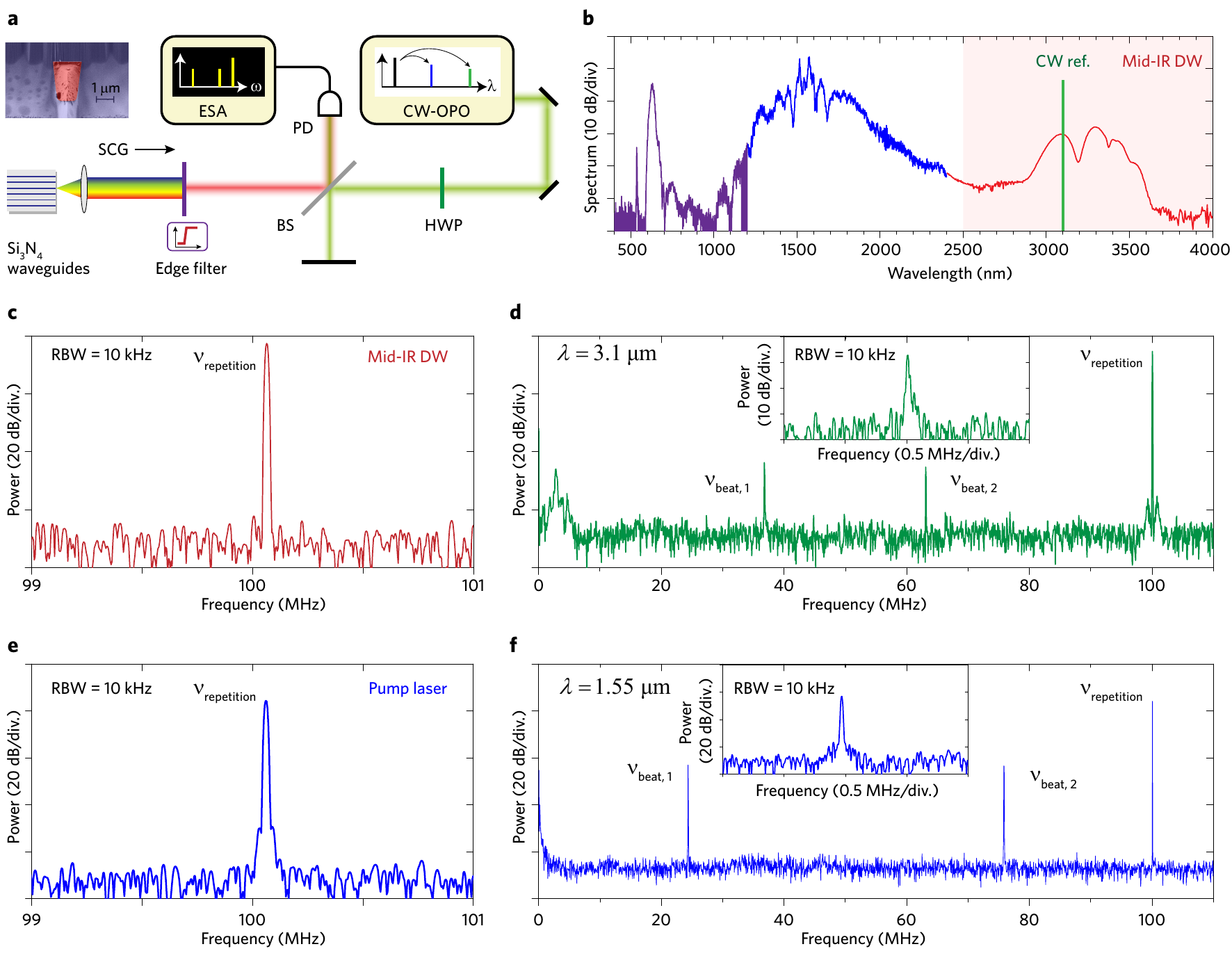}
  \caption{\label{fig-4} \textbf{Coherence measurements of the mid-IR frequency comb} (a) Sketch of the experimental setup for characterizing both the repetition beatnote of the mid-IR wave alone, and a heterodyne beatnote with a continuous wave reference laser in the mid-IR, via free-space optical alignment. ESA: electrical spectrum analyzer; CW-OPO: continuous wave optical parametric oscillator; PD: photodetector; BS: beam splitter; HWP: (mid-IR) half-wave plate. (b) Spectrum of the generated supercontinuum in which the mid-IR wave is filtered by an long-pass edge-filter (red shading area, cut-on wavelength at ${2.5~\mu{\rm m}}$), the filtered mid-IR wave has an overall power of ${\sim 100 ~\mu{\rm W}}$. (c) The repetition beatnote of the mid-IR wave (resolution bandwidth ${10 ~{\rm kHz}}$), signal-to-noise ratio greater than ${70 ~{\rm dB}}$, linewidth is ${\sim 50 ~{\rm kHz}}$. (d) The heterodyne beatnote with the mid-IR reference laser (wavelength at ${\sim 3.1 ~\mu{\rm m}}$, power ${\sim 10 ~{\rm mW}}$), in which a pair of beatnotes are detected within one free spectral range (${\sim 100 ~{\rm MHz}}$). Inset shows a zoomed-in CW-beatnote that has a signal-to-noise ratio greater than ${20 ~{\rm dB}}$ and a linewidth below ${50 ~{\rm kHz}}$ (resolution bandwidth ${10 ~{\rm kHz}}$). (e,f) repetition beatnote and CW-beatnote (using a low-noise single-frequency fiber laser) of the pump laser.}
  }
\end{figure*}

A numerical simulation based on the nonlinear Schr\"odinger equation (NLSE) could reproduce the supercontinuum with a high level of agreement, see Fig. \ref{fig-2} (b), which further provides an insight into soliton dynamics during the propagation.
First, the first compression point is around ${3 ~{\rm mm}}$ where dispersive waves start to be emitted, see Fig. \ref{fig-2} (c), which matches with experimental observations of visible green light scattered off the waveguide at the corresponding position.
To simulate a close-to-realistic pump source, we pre-chirped the pulse at the input to match a comparable duration. This leads to almost no broadening in the beginning stage of the propagation.
As a consequence, the first compression point occurs at a longer propagation distance than what is estimated.
Second, the wavelengths of both visible and mid-IR dispersive waves match well with experimental observations. The phase matching condition is confirmed at these wavelengths by the waveguide dispersion simulation.
With dispersive waves on both wavelength sides of the pump pulse, the soliton recoil effect is mutually compensated, resulting in negligible spectral shift of the soliton.
Moreover, ripples on the spectrum can be understood as the result of the soliton fission effect by which the input pulse is split into several solitons that lead to a spectral interference. The number of split solitons is determined by the soliton number ${N}$.
With a low soliton number, the soliton-fission-induced incoherence is weak.
Therefore, the supercontinuum with soliton-seeded dispersive waves is likely to inherit the high coherence of the mode-locked pump laser.
In simulations, the degree of first-order coherence ${g_{12}^{(1)}}$ can be evaluated via \cite{Dudley2}:
\begin{equation}
{g_{12}^{(1)}(\lambda )} = {\frac{{\left\langle {E_l^*(\lambda ){E_s}(\lambda )} \right\rangle }}{{\sqrt {\left\langle {|{E_l}(\lambda )|^2} \right\rangle \left\langle {|{E_s}(\lambda )|^2} \right\rangle } }}}, ~ s \ne l
\end{equation}
where ${E_l}$ and ${E_s}$ are the envelopes of supercontinua from two independent simulations with randomly generated noise, equivalent to one photon per mode.
An averaged modulus of ${\left| {g_{12}^{(1)}} \right|}$ considering 50 runs of the simulation indicates that the presented spectrum is robust to the noise, see Fig. \ref{fig-2} (d).

Moreover, comparing supercontinua from waveguides with different widths, we observed the tuning of the mid-IR dispersive wave, see Fig. \ref{fig-3}(a,b), as the phase-matching wavelength is modified.
Interestingly, in our ${\rm Si_3N_4}$ waveguides, we notice a double benefit from increasing the waveguide width.
First, the enlarged cross section favors light guiding at longer wavelengths in mid-IR.
Second, the dispersive wave is generated further in the mid-IR as well.
However, this set of waveguides shows weak confinement at around ${4 ~\mu{\rm m}}$, mostly limited by the height (${0.87 ~\mu{\rm m}}$)-- a feature dictated by the ${\rm Si_3N_4}$ deposition process.
It is also noted that the measured supercontinuum features water absorption lines in the range of ${2.5-2.9 ~\mu{\rm m}}$, which is attributed to occur within the FTIR, see Fig. \ref{fig-3}(b).


To achieve mid-IR dispersive wave generation at even longer wavelengths, we turn to large-cross-section ${\rm Si_3N_4}$ waveguides which have a height of ${2.3~\mu{\rm m}}$ and a width of ${0.9-1.1 ~\mu{\rm m}}$.
Such waveguides then enable wave confinement for longer wavelengths and a larger flexibility of dispersion engineering, compared to the first class.
The simulated dispersion profile indicates that in such waveguides the dispersive wave phase matching wavelength can be tuned over the wavelength range ${2.5-4 ~\mu{\rm m}}$, see Fig. \ref{fig-3} (c), which covers the functional group region.
Indeed, when pumping with the telecom-band femtosecond laser, we observed ultra-broadband SCG in combination with mid-IR dispersive wave generation in this range, see Fig. \ref{fig-3} (d).
Apart from the mid-IR dispersive wave, the wavelength of the visible dispersive wave (generated simultaneously in the waveguide) is slightly changed, as indicated by the phase matching condition.
Raised by two dispersive waves, the supercontinuum in the large-cross-section waveguides can span up to \emph{three} octaves.
In spite of a conversion efficiency from the pump to the mid-IR dispersive wave that decreases when the latter red-shifts,  the low soliton number configuration should lead to a high coherence degree over most of the generated supercontinua.

\section{Coherence measurement of the Mid-infrared frequency comb }
\label{sec:comb}

Ideally, the mid-IR dispersive wave would perform as a frequency comb that inherits the coherence from the pump source. However, it is critical to confirm the coherence, which can be lost due to a variety of processes.
Here, we experimentally investigate the phase coherence of such mid-IR frequency combs generated in our ${\rm Si_3N_4}$ waveguides by beatnote measurements with a narrow-linewidth continuous wave (CW).
Such a characterization technique for assessing comb coherence properties is known as an alternative \cite{cundiff2003colloquium,gohle2005frequency,wang2013mid,burghoff2014terahertz,kuyken2015octave} to that involving the ${f - 2f}$ interferometer \cite{udem2002optical,cundiff2003colloquium}.
In the characterization, the mid-IR wave packet of the supercontinuum is filtered by a long-pass edge-filter (wavelength at ${2.5 ~\mu{\rm m}}$). We subsequently both measure the repetition beatnote of the filtered mid-IR frequency comb alone, and beat it with a tunable CW-OPO (Argos Aculight, ${2.5-3.2 ~\mu{\rm m}}$, linewidth ${<1 ~{\rm MHz}}$), on a fast IR detector (Vigo, PVMI-4TE-8), see Fig. \ref{fig-4}(a) for a sketch of the setup.

Figure \ref{fig-4}(b) shows one optical spectrum of the supercontinuum used for the mid-IR coherence measurement, which is generated in a waveguide with the cross-section of ${1.0 \times 2.3 ~\mu{\rm m^2}}$. Such a supercontinuum contains a mid-IR wave packet that is well separated from the spectral contents in the visible and the near-IR, spanning from ${2.8-3.6 ~\mu{\rm m}}$, and maintains its spectral envelope after the edge-filter. The overall power in the mid-IR is ${\sim 100 ~\mu{\rm W}}$. The repetition beatnote of such a mid-IR frequency comb is shown in Fig. \ref{fig-4}(c), under a resolution bandwidth of ${10 ~{\rm kHz}}$, where a narrow-linewidth beat signal is observed indicating the repetition frequency (${\sim 100 ~{\rm MHz}}$) of the comb. The signal-to-noise (SNR) ratio is above ${70 ~{\rm dB}}$ and the short-term linewidth is ${\sim 50 ~{\rm kHz}}$. The heterodyne beatnote between the mid-IR frequency comb and the CW-OPO, at the wavelength of ${3.1 ~\mu{\rm m}}$, is shown in \ref{fig-4}(d), where within a span of ${110 ~{\rm MHz}}$ three isolated beatnotes are observed. The strongest beat signal at ${\sim 100 ~{\rm MHz}}$ again corresponds to the repetition frequency, while the other two beatnotes correspond to the beat signal generated by the CW-OPO and the two spectrally closest lines of the mid-IR frequency comb. Such CW-beatnotes have a SNR ratio above ${20 ~{\rm dB}}$ and a linewidth below ${50 ~{\rm kHz}}$.
We also directly characterized the femtosecond seed laser in the telecom-band, in terms of measuring both the repetition beatnote (Fig. \ref{fig-4}(e)) and the CW-beatnote (Fig. \ref{fig-4}(f)) with a low-noise single-frequency fiber laser (NKT Koheras, linewidth ${< 100 ~{\rm kHz}}$). Compared with the coherence properties of the seed laser, no noticeable broadening is observed in both types of beatnote for the mid-IR frequency comb, implying a high-level of coherence inherited from the seed laser in the telecom-band to the mid-IR wave packet, via SCG. Our investigation therefore demonstrates a coherent regime of mid-IR frequency comb generation.

\section{Conclusion}
\label{sec:conclu}

We presented a compact and simple platform for mid-IR optical frequency comb generation which can access the functional group region ${2.5-4 ~\mu{\rm m}}$ through dispersive wave generation in the coherent supercontinuum process. This process allows for pumping in the telecom-band where a variety of seed lasers, e.g. erbium-fiber based femtosecond laser, can be accessed.
Using large-cross-section ${\rm Si_3N_4}$ nano-photonic waveguides, spectral broadening is governed by both the SPM effect in the soliton self-compression regime, and dispersive wave formation in both the visible and the mid-IR range, resulting in an overall bandwidth spanning up to three octaves. In particular, the high-level phase coherence can be transferred from the seed laser to the dispersive wave in the mid-IR.
The ability to achieve a mid-IR optical frequency comb directly from a telecom-band femtosecond laser with a chip-based nano-photonic waveguide has several promising applications.
In contrast to other approaches, based on OPO or DFG for example, the present approach can be very cost effective and compact, and can be readily extended for dual-comb spectroscopy \cite{lee2001ultrahigh, schiller2002spectrometry, keilmann2004time, schliesser2005frequency, yasui2005asynchronous}.

\begin{acknowledgments}
	This publication was supported by Contract W31P4Q-16-1-0002 (SCOUT) from the Defense Advanced Research Projects Agency (DARPA), Defense Sciences Office (DSO). This material is based upon work supported by the Air Force Office of Scientific Research, Air Force Material Command, USAF under Award No. FA9550-15-1-0099. H.G. acknowledges support by funding from the European Union’s Horizon 2020 research and innovation programme under Marie Sklodowska-Curie IF grant agreement No. 709249. A.B., D.G. and C.S.B. acknowledge support from the European Research Council under grant agreement ERC-2012- StG 306630-MATISSE. All samples were fabricated and grown in the Center of MicroNanoTechnology (CMi) at EPFL.
\end{acknowledgments}

\section*{Methods}
\label{sec:method}

\begin{figure}[tbh]
  \centering{
  \includegraphics{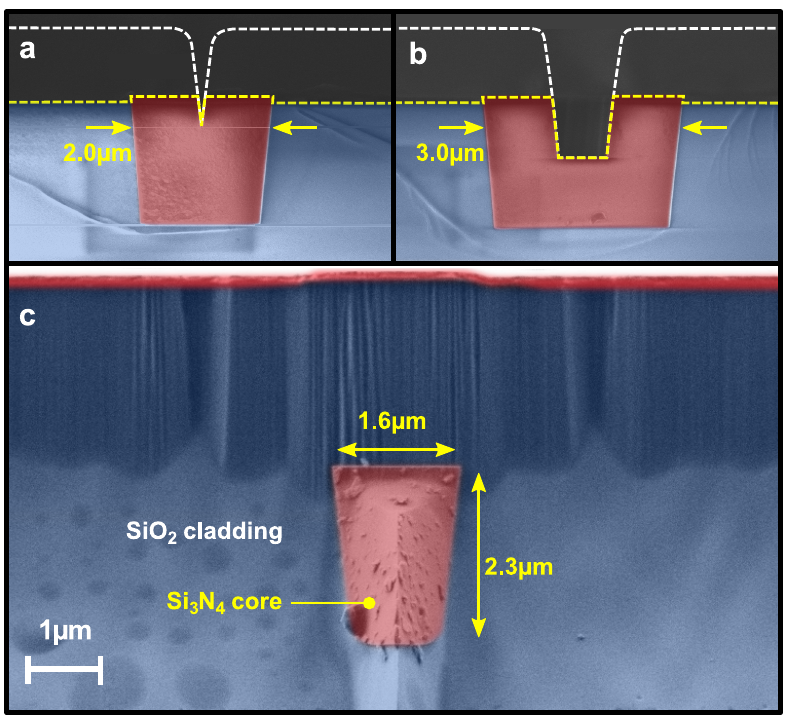}
  \caption{\label{fig-5} \textbf{Cross sections showing the thick waveguide fabrication scheme.} (a) and (b) show cross sections imaged with SEM after the surface planarization. The dashed white line indicates the cross section after ${\rm Si_3N_4}$ deposition. From (b) it can be seen that the gap is filled conformally during the deposition, i.e. the film is growing from the bottom as well as from the sides of the trench. The transition to a completely filled gap is shown in (a), as the trench width is reduced to ${\leq 2}$ times the ${\rm Si_3N_4}$ film thickness. A fabricated waveguide with a height of ${2.3 ~\mu{\rm m}}$ and a width of ${1.6 ~\mu{\rm m}}$  is shown in (c).}
  }
\end{figure}

\noindent \textbf{Large-size ${\rm \bf Si_3N_4}$ waveguides beyond cracking limitation} ---
The fabrication of large-size ${\rm Si_3N_4}$ waveguides is challenging as the tensile stress in ${\rm Si_3N_4}$ layers by LPCVD can lead to cracks when the layer thickness is increased.
In practice, the maximum crack-free ${\rm Si_3N_4}$ film thickness for our process was found to be $\sim 1.0~\mu{\rm m}$.
Here we achieve the fabrication of a large-size ${\rm Si_3N_4}$ waveguides with a height as large as ${2.3~\mu{\rm m}}$ using the photonic Damascene process.
Importantly, we make use of the high surface mobility provided by the dichlorosilane (SiH$_2$Cl$_2$) and ammonia (NH$_3$) chemistry during LPCVD.
This allows a high conformality when depositing the ${\rm Si_3N_4}$ film onto the pre-structured substrate.
The result is shown in Fig. \ref{fig-5} on cross sections of filled trenches with different widths.
In the panels (a) and (b), the cross sections were imaged after the planarization step, included in the photonic Damascene process flow.
The dashed white line indicates the cross section of the ${\rm Si_3N_4}$ film right after deposition.
From Fig. \ref{fig-5} (b) it can be seen that the ${\rm SiO_2}$ preform (blue) is conformally filled with ${\rm Si_3N_4}$ (red), i.e. deposition rates  for layers growing from the bottom and from the sides are identical.
As the trench width approaches less than 2-times the film thickness (i.e. $2 \times 1.0~\mu{\rm m}$), the gap begins to close, as shown in Fig. \ref{fig-5} (a).
This enables us to fabricate waveguides with a height of ${2.3~\mu{\rm m}}$ and a maximum width of ${1.8~\mu{\rm m}}$.
No voids were observed in such large-size waveguides.
There is a slightly slanted sidewall angle of ${\sim 85^\circ}$.
An example of such a waveguide with a width of $1.6~\mu{\rm m}$ is shown in Fig. \ref{fig-5} (c).

\end{document}